# Transmission behaviors of single mode hollow metallic waveguides dedicated to mid-infrared nulling interferometry

Laëtitia Abel-Tibérini[1*], Lucas Labadie[1,2], Brahim Arezki[1], Pierre Kern[1], Romain Grille[1], Pierre Labeye[3] and Jean-Emmanuel Broquin[4]

[1] Laboratoire d'Astrophysique Observatoire de Grenoble, BP53, F-38041 Grenoble Cedex 9 (France)
[2] Max-Planck Institut für Astronomie, Königstuhl 17, D-69117 Heidelberg (Germany)
[3] LETI/DIHS/LMNO, 17 rue des Martyrs, BP 85X, 38054 Grenoble Cedex 9 (France)
[4] Institut de Microélectronique, Electromagnétisme et Photonique, MINATEC-INPG, 3 Parvis Louis Néel, BP 257, 38016 Grenoble Cedex 1 (France)

*Corresponding author: laetitia.abel@fresnel.fr*

**Abstract:** This paper reports the characterization of hollow metallic waveguides (HMW) to be used as single-mode wavefront filters for nulling interferometry in the 6-20µm range. The measurements presented here were performed using both single-mode and multimode conductive waveguides at 10.6µm. We found propagation losses of about 16dB/mm, which are mainly due to the theoretical skin effect absorption in addition to the roughness of the waveguide's metallic walls. The input and output coupling efficiency of our samples has been improved by adding tapers to minimize the impedance mismatch. A proper distinction between propagation losses and coupling losses is presented. Despite their elevate propagation losses, HMW show excellent spatial filtering capabilities in a spectral range where photonics technologies are only emerging.

## 1. Introduction

In the context of the search for new extrasolar systems, the direct detection of Earth-like planets around main sequence stars is facing serious constraints both in terms of angular resolution ($< 0.1$ arcsec) and brightness contrast ($>10^6$) between the star and the planetary companion. Nulling interferometry [1] is a coronographic technique at very high angular resolution, particularly adapted to the search and characterization of extrasolar planets in the mid-infrared range (6-20µm), where the relative Earth/Sun contrast is more favorable. The principle of a two aperture nulling interferometer is to recombine artificially $\pi$ phase-shifted wavefronts from an on-axis star in order to achieve their mutual cancellation, while the signal from an off-axis planet is transmitted by tuning the interferometer baseline so that a total phase shift of $2\pi$ is achieved. This technique is foreseen as the core of the future planets haunting mission Darwin/TPF [2]. As in classical co-axial stellar interferometer, the rejection ratio of a nuller can be sensitively degraded by the effect of low-order and high-order wavefront errors, which can be efficiently filtered out by mean of single-mode optical waveguides [3]. In addition to their filtering properties, single mode waveguides can be integrated in stable, light and compact optical chips to ensure more complex functions like multiple beams combination [4]. Integrated optics devices are well-known in the near-infrared range and commonly used for stellar interferometry in the near infrared [5]. In the mid infrared range, the availability of single-mode waveguides is only recently emerging, making their development a strong challenge in the field of photonics [6, 7, 8]. An alternative to dielectric waveguides – based on silver-halide or chalcogenide materials – are the *conductive waveguides* or HMW, which are similar to the microwave metallic waveguides but scaled down to the mid-infrared regime [9, 10]. We present the measured losses in different conductive waveguides and discriminate between their potential origin (coupling effects, propagation losses). We complete our analysis by the characterization of bent waveguides and T-junctions. Near-field output imaging is also consider here as a diagnostic tool of the waveguides modal behavior.

## 2. Design, manufacturing and characterization procedure of rectangular conductive waveguides

### 2.1. Building the elementary blocks

The samples characterized in this paper are micrometric rectangular waveguides obtained using a standard micro-technology etching process of silicon followed by the anodic bonding of a pyrex superstrate [10]. After the etching process, the waveguide walls are coated with gold since this metal offers a high reflectivity at 10.6µm and a sufficiently low surface roughness of 50nm rms [11]. Fig. 1(a) shows the geometry of a conductive waveguide. The electromagnetic wave propagates in the air by successive bouncing on the gold-coated walls. Its dimensions *b* and *a* are the only parameters influencing the modal behavior of the structure as a function of the wavelength. If $b<a$, the spectral range where only the fundamental mode $TE_{10}$ can propagate is given by $2b<\lambda<2a$ [12]. Playing on these geometrical parameters during the manufacturing process allows tuning the single-mode spectral range. A good compromise between size issues and single-mode behavior at $\lambda=10.6\mu m$ is a waveguide with dimensions $a=10\mu m$ and $b=5\mu m$. At the same wavelength, a waveguide with $a=10\mu m$ and $b=7.5\mu m$ is multimode by design. Fig. 1(b) presents a cross section of a 10x5µm structure obtained with a Scanning Electron Microscope. For the purpose of the analyze, the manufactured waveguides have their *a* dimension that stays fixed to 10µm while their *b* dimension ranges from 4.5 to 10µm, allowing parametric analysis of the modal behavior according to waveguide section. Fig. 1(c) is a picture of a chip containing several HMW with different sizes.

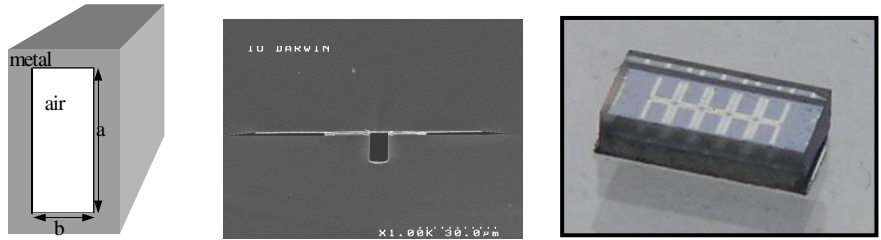

Fig. 1: (a) Hollow Metallic Waveguide: geometry (b) SEM images of the output of a 10x5µm guide, (c) Photograph of a chip containing several waveguides. The chip is 1-mm high, 1-mm wide and 5-mm long.

The numerical aperture imposed by the waveguide geometry is very constraining (f/0.4). To reduce this difficulty, additional tapers are integrated to the waveguide input and output to enlarge the section from 10x5µm to 10x40µm over a 100µm length. The numerical aperture of the incoming beam to be coupled to the waveguide input is enlarged to f/1.15, which remains an achievable value for standard optics [13].

*2.2. Laboratory setup and measurement procedure*

In a previous paper [13], we obtained the first characterization of single-mode conductive waveguides at 10.6µm. Since then, new samples with different lengths have been manufactured in order to allow a more accurate estimation of the propagation losses based on differential measurements. All characterizations presented in this paper were performed at 10.6µm using a $CO_2$ laser. Using the optical set-up presented in Fig. 2, the laser beam is focused onto a 50µm pinhole. The resulting unresolved object is re-imaged on the waveguide input using an aspherical f/1.15 injection lens in order to match the tapered aperture. The excellent optical quality of this lens results in diffraction limited injection spot with 25µm full-width at half maximum. A quarter-wave plate ensures a circular polarization of the incoming beam, which allows considering all the polarization directions as we characterize both single-mode and multimode waveguides. A set of two lenses is used to image the output of the HMW on a bolometric camera with a magnification $\gamma=6$. The first imaging lens has an f/0.28 numerical aperture, which permits to spatially resolve the electric field distribution at the taper output. The dynamic range of the detection means is 37dB including the dynamic of the camera and an additional set of neutral densities. The measurements of transmissions lower than 0.01 % was not attainable using this bench. Such measurements should require performing the detection of fainter fluxes with respect to the environment. The implementation of a cooling system for the overall setup would be one of the solutions to improve significantly its performances at the considered wavelengths.

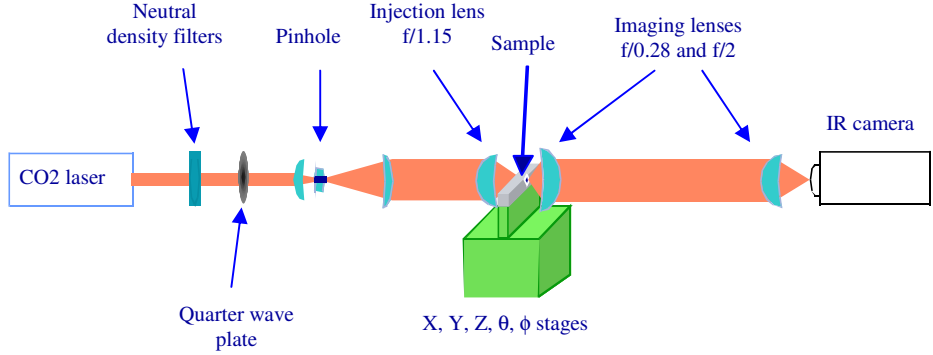

Fig. 2: Optical characterization set-up for transmission measurement of conductive waveguides.

The principle of the characterization is the following. The first image records $\Phi_{HMW}$, the transmitted flux with a HMW in place on the bench. The second image records $\Phi_{ref}$, the transmitted flux without any component on the bench. A calibration step is performed to take into account the response of the detector to a flat field illumination as well as the dark current of the detector array. The global transmission of the component is then given by the direct ratio of the calibrated fluxes [13]:

$$T = \frac{\Phi_{C,in}}{\Phi_{C,out}} \qquad (1)$$

where $\Phi_{C,\,in}$ is the calibrated flux through the component and $\Phi_{C,\,out}$ the calibrated flux with no component installed.

### 3. Results on the characterization of the hollow metallic waveguides

The optical bench described above provides a measurement of the transmission for the whole component, including coupling and propagation losses. The discrimination between the origins of the losses of the component is achieved by a direct comparison of the transmission measurements obtained for different configurations of the waveguide.

#### 3.1. Estimation of the propagation and coupling losses

The propagation losses are estimated by comparing waveguides with identical cross section but with different lengths. The subtraction of the two measured transmissions removes the coupling contribution. This is represented in Fig. 3 by the solid line vertical arrow. The effect of the coupling on the global transmission is estimated by comparing transmission measurements on waveguides with identical length, but *with* and *without* tapers. This is represented in Fig. 3 by the dashed line horizontal arrow. The bold-faced transmission values in the boxes are the *experimental* values. The transmission values in simple characters next to the arrows are the derived values. The values presented in Fig. 3 are for the particular case of waveguides with a 10x5µm cross section.

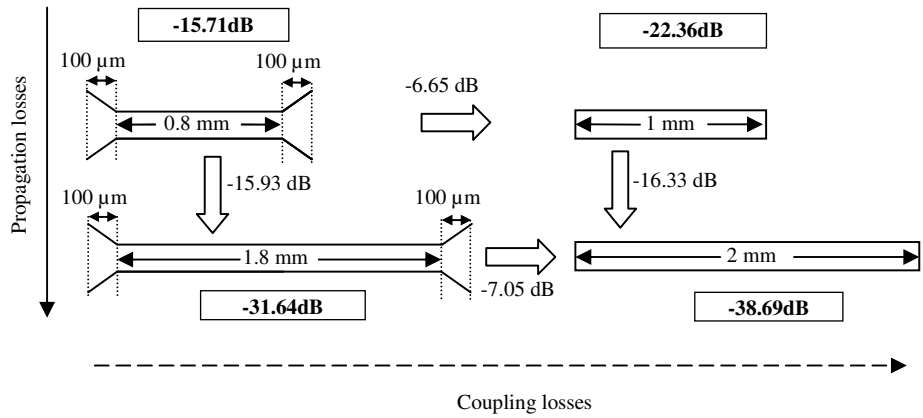

Fig. 3: Comparison of measured transmissions between HMW with identical cross sections (10 x 5 µm). In the vertical direction, the propagation losses are estimated by comparing transmissions between waveguides with different lengths (0.8mm versus 1.8mm and 1mm versus 2mm). An average value of 16dB/mm propagation losses is derived from the experimental data. In the horizontal direction, we estimate the effect of the coupling taper on the global transmission.

This procedure was conducted over a large number of components in order to gain a statistical measurement. In this way, we have also estimated the dispersion introduced by the differences in the optical quality of the input and output facets of the chips, due mainly to the process of wafer cutting and dicing. Two distinct measurement campaigns were conducted for single-mode waveguides having respectively a cross section geometry of 10x5µm and 7.5x4.5µm. Table 1 reports the results for the global transmissions T(1-mm) and T(2-mm), as well as the derived propagation and coupling losses. We underline that we considered the effective propagation length of 0.8mm for the tapered waveguides.

The standard deviation (i.e. the absolute error) on the measured transmissions is 0.02%. Integrating a taper to our waveguides shows a significant improvement by almost 7dB of the global transmission for the samples with a 10x5µm cross section. It was unfortunately not possible to carry out an equivalent measurement on the samples with a 7.5x4.5µm cross section. At this stage, it is difficult to know what is the effect of a single taper since the obtained value cannot be simply divided by two because the numerical aperture of the input and output lenses of the setup are different.

Table 1: Measurement of the transmission of different conductive waveguides with cross sections 10x5µm and 7.5x4.5µm.

| | | T(1mm) | T(2mm) | α | Average coupling improvement |
|---|---|---|---|---|---|
| Waveguide cross section | 10x5µm | 2.69 % (~ -15.7 dB) | 0.07 % (~ -31.5 dB) | 2.60 % (~ -15.8 dB/mm) | + 6.85 dB |
| | 7.5x4.5µm | 0.40% (~ -24 dB) | 0.01% (~ -40 dB) | 2.50% (~ -16.02 dB/mm) | N.A. |

The last column presents the improvement factor in dB obtained by adding a taper at the input and the output of the waveguides. The α factor gives the propagation losses.

In terms of propagation losses, the experimental results show similar values of 16dB/mm for both types of structures, which is considerably higher than what estimated in our previous work [13]. In addition, the experimental propagation losses seem to be less affected by the waveguide cross section than what was predicted theoretically [11]. Considering the error bars of our measurements, the *skin effect* model applied to the fundamental mode $TE_{10}$ of a metallic waveguides [11] predicts theoretically similar losses 15.5dB/mm for the 7.5x4.5µm geometry, while it predicts a higher transmission with 10.5dB/mm losses for the 10x5µm geometry. Other potential causes of losses can be considered to explain the obtained values. The 50nm rms roughness of the metallic walls is larger than the skin depth of $\delta = 13.5$nm and, according to Holloway et al. [14], this generates additional losses of ~1.5dB. Furthermore, we have used a circularly polarized radiation to characterize our components. Since a single-mode conductive waveguide works like an integrated linear polarizer, only half of the power corresponding to the matching polarization direction is transmitted through the component.

*3.2. Towards more complex optical functions: bent waveguides and T-junctions*

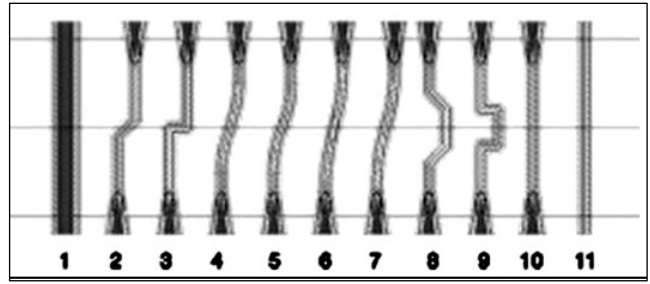

Fig. 4: Design of a chip containing waveguides with 45°, 90°, double 45°, double 90° angles (#2, #3, #8, #9). Other waveguides are S-bent with constant or continuous radii of curvature from 500 µm to 600 µm (#4, #5, #6, #7). On the left, a larger waveguide (100 µm) simplifies the alignment of the chip. The two straight waveguides on the right (with and without taper) gives the reference measurement for the transmission value of the chip.

For sake of completeness of our study, we have evaluated the losses observed in waveguides with curvatures, as precursors to more complex functions. Several chips containing single-mode and multimode bent waveguides with different radius of curvature were manufactured in a second phase of the project (Fig. 4). The transmission was estimated for each waveguide using the same procedure than above. The results are reported in the graph of Fig. 5.

We found that the waveguides with a constant or continuous radius of curvature (#4, #5, #6, #7) present a better transmission that waveguides with more abrupt 45° or 90° angles (#2, #3). A 45° angle induces less losses than a 90° one (which transmission is found below the noise level) and, obviously, two successive angles (#8, #9) induce more losses than in the previous case. We also note that the waveguides with a radius of curvature of R = 600µm (#5 and #7) transmit slightly better than the ones with R = 500µm (#4 and #6). These results confirm that it would be preferable to use bent waveguides with large radii of curvature for the operation of beam combination. Quantitatively, it has been found that a waveguide with a 500µm radius of curvature introduces a 1.2dB additional loss.

Integrated beam splitters were characterized on our bench to estimate the effect of the splitting function on the global transmission. Fig. 6 presents the layout of the tested T-junctions.

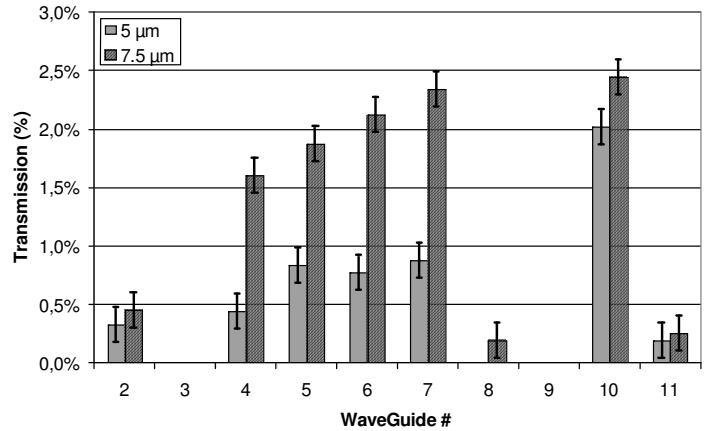

Fig. 5: Transmission for different waveguides with 10x5 µm section (gray columns) and 10x7.5µm sections (hatched columns). The waveguide numbers are the same than for Fig. 4. The missing values (waveguides #3 and #9), correspond to measurements with a too low signal to noise ratio.

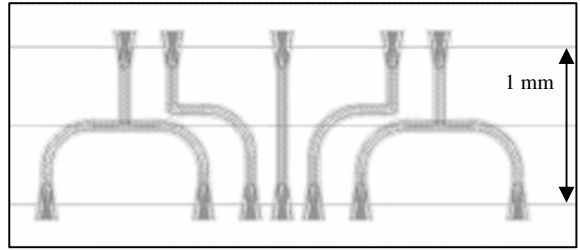

Fig. 6: A 1-mm long chip with 2 identical T junctions (on the right and the left). For the T factor characterization the incident beam is injected from the top.

At this stage no significant transmitted flux was detected for several T-junctions with a cross section of 10x7.5µm. Several reasons account for a poor transmission of these components: the splitting function induces a 3dB additional loss; the waveguide radius of curvature being 300µm, the resulting loss is higher than the 1.2dB in subsection 3.2.; the effective length of the T-junction is 400µm longer than the channel straight waveguides of 1mm long characterized above, which results in loss increase of 6.4dB considering the values for the propagation losses reported in Table 1. The resulting overall losses lead to a transmitted flux through the component which remains lower than our detection threshold.

Detections of transmitted flux was achieved only using multimode T- junction with a cross section of 10x10µm. The very poor signal to noise of the observed flux did not allow any further interpretation of these results.

## 4. Near-field imaging to characterize the modal behavior

Several settings have been proposed to characterize the single mode behavior of optical waveguides. In a previous paper [13], we proposed a characterization method based on the measurement of the polarization state of the flux transmitted through a conductive waveguide that was successfully applied to our samples. This method is limited by the extinction ratio of the polarization device. A second approach is based on the measurement of interferometric contrasts [15] in a nulling interferometer as a direct application of single-mode waveguides for spatial filtering. Here, we adopt an approach based on the spatially resolved imaging of the waveguide output field. We tested this method on waveguides expected to be, by design, single-mode (10x5µm) or multimode (10x10µm) at λ=10.6µm. The tested components have a 10x40µm taper on each extremity.

The cross section of the field is recorded at the taper output along its 40µm edge. The imaging system is used to insure a magnification of 25 allowing sampling the 40µm field with 25 pixels. The f/0.28 numerical aperture of the first lens provides a spatial resolution of 3µm at λ=10.6µm. We assume here that the output taper is perfect in the sense that it does not introduce any additional propagation modes than the ones transmitted by the channel waveguide itself as described in Fig. 7.

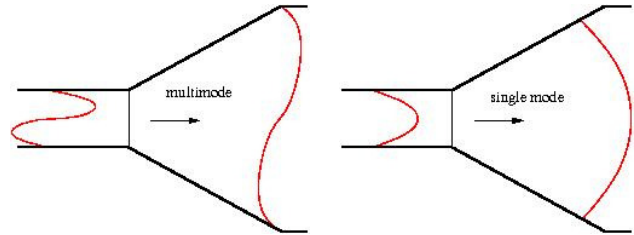

Fig. 7: Schematic view of the behavior of the output taper with respect to a propagation mode. The taper ensures the confinement of the electric field without introducing any additional mode.

The theoretical two-dimensions field distribution over the waveguide aperture as a function of the cross section geometry and the wavelength is described in details in [13].

It is experimentaly observed that the field distribution over this cross section is not affected by the displacement of the injection beam in the case of the single mode waveguide, as it is strongly affected in the case of the multimode waveguide as it shown in Fig.8. The only affected behaviour is the transmited flux in the case of the single mode waveguide.

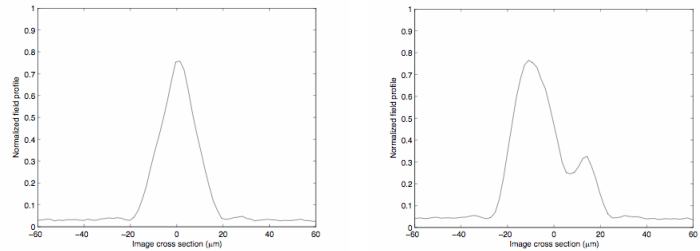

Fig. 8: Cross section of the image of the output field for a single mode (10x4.5µm) waveguide (left), and for a multimode (10x10µm) waveguide (right). In both cases, a 10x40µm taper is implemented on each guide extremity.

## 5. Discussion and conclusions

We presented in this paper the characterization at λ=10.6μm of conductive waveguides with different geometries, including channel waveguides, bent waveguides and beam splitters. We found that the propagation losses of these structures are generally high, i.e. 16dB/mm. Since these components are designed to work primarily as spatial filters, a compromise between propagation losses and wavefront filtering capabilities has to be searched playing on the total length of the waveguide. Let us consider the attenuation coefficient $\alpha_{HM}$ in dB/m of the high order modes in a single-mode waveguide given by [13]:

$$\alpha_{HM} = 8.68 \times \left(\frac{2\pi}{\lambda}\right)\left[\left(\frac{\lambda}{\lambda_C}\right)^2 - 1\right]^{1/2} \quad (2)$$

where λ is the wavelength, $\lambda_c$ is the cut-off wavelength of the considered mode. A rectangular conductive waveguide with $a = 2b = 10$μm is single mode for 10μm<λ<20μm. The cut-off wavelength of the first higher order modes – $TE_{20}$ and $TE_{01}$ – is $\lambda_c = 10$μm. At λ=10.6μm, the resulting attenuation is $\alpha_{HM}(TE_{20}) = \alpha_{HM}(TE_{01}) = 1.8$dB/μm. Thus to reach an attenuation of the high order modes equal to $10^6$ as required for nulling interferometry – which is equivalent to -60 dB – a waveguide length of 40μm is sufficient. The resulting losses of the fundamental mode due to propagation give a waveguide transmission, derived from the measurements presented above, of approximately 0.64dB, or 85%.

## Acknowledgements

This work was funded by the European Space Agency contract 16847/02/NL/SFe, the French Space Agency (CNES) and Alcatel-Alenia Space.